\documentclass[twocolumn,showpacs,aps,prl,superscriptaddress,unsortedaddress]{revtex4}
\usepackage{graphicx}
\usepackage{epsf}
\usepackage{hyperref}

\begin{document}
	
\title{Optical Birefringence and Dichroism of Cuprate Superconductors in the THz regime}
\author{Y. Lubashevsky}
\affiliation{The Institute of Quantum Matter, Department of Physics and Astronomy, Johns Hopkins University, Baltimore, MD 21218, USA}
\author{LiDong Pan}
\affiliation{The Institute of Quantum Matter, Department of Physics and Astronomy, Johns Hopkins University, Baltimore, MD 21218, USA}
\author{T. Kirzhner}
\thanks{Deceased.}
\affiliation{Department of Physics, Technion, Haifa 32000, Israel}
\author{G. Koren}
\affiliation{Department of Physics, Technion, Haifa 32000, Israel}
\author{N. P. Armitage}
\affiliation{The Institute of Quantum Matter, Department of Physics and Astronomy, Johns Hopkins University, Baltimore, MD 21218, USA}

\begin{abstract}
The presence of optical polarization anisotropies, such as Faraday/Kerr effects, linear birefringence, and magnetoelectric birefringence are evidence for broken symmetry states of matter.  The recent discovery of a Kerr effect using near-IR light in the pseudogap phase of the cuprates can be regarded as a strong evidence for a spontaneous symmetry breaking and the existence of an anomalous long-range ordered state. In this work we present a high precision study of the polarimetry properties of the cuprates in the THz regime. While no Faraday effect was found in this frequency range to the limits of our experimental uncertainty (1.3 milli-radian or 0.07$^\circ$), a small but significant polarization rotation was detected that derives from an anomalous linear dichroism.  In  YBa$_2$Cu$_3$O$_y$ the effect has a temperature onset that mirrors the pseudogap temperature T$^*$ and is enhanced in magnitude in underdoped samples.    In $x=1/8$  La$_{2-x}$Ba$_{x}$CuO$_4$, the effect onsets above room temperature, but shows a dramatic enhancement near a temperature scale known to be associated with spin and charge ordered states.  These features are consistent with a loss of both C$_4$ rotation and mirror symmetry in the electronic structure of the CuO$_2$ planes in the pseudogap state.   
\end{abstract}

\date{\today}

\maketitle

An extensive research effort has been carried out over the last two decades on defining the role and origin of the pseudogap phase in the cuprates. The pseudogap, a regime of the phase diagram generally located at higher temperatures than the superconducting state, is characterized by an energy gap in the density of states at the Fermi level as well as various transport and magnetic anomalies. Whether this gap is related to superconductivity or competes with it, and whether it realizes an additional long-range ordered state is controversial\cite{Timusk,Norman}.  Characterization of a stable static order with true broken symmetry in the pseudogap regime could solve the mystery surrounding its origin.

Optical polarization anisotropies, such as Faraday/Kerr effects, gyrotropic rotation, linear birefringence, and magneto-electric birefringence can be sensitive tools for the detection of broken symmetry states of matter.  For instance, materials with anti-symmetric off-diagonal components in the dielectric tensor can rotate the plane of polarization of linearly polarized light.   Such tensor elements are only allowed in a material that breaks either time-reversal or inversion and mirror symmetries.  Such effects are referred to as circular, since the eigenmodes of their transmission or reflection matrices are left and right circular polarizations.  The most common such ``circular" effects are magneto-optical ones arising from time-reversal symmetry breaking from magnetic moments aligned either by applying external magnetic field or spontaneous magnetization.  Another circular effect arises in so-called gyrotropic ordered materials that breaks all mirror symmetries.  Spiral structures and cholesteric textures have such optical activity and can rotate polarization in the absence of magnetic moments.    Linear dichroism or birefringence in, say, an orthorhombic crystal manifests the breaking of  C$_4$ rotation symmetry.  Polarization anisotropies can be further characterized by them being $reciprocal$ or not.  Effects are non-reciprocal if they have opposite signs for the two states of a crystal that are related to each other by time reversal.  In certain circumstances non-reciprocity can be probed effectively by considering propagation through the backside of a sample in the reverse direction or by ``flipping" it.   Non-reciprocity is a sufficient, but not necessary condition for microscopic time-reversal symmetry breaking \cite{de-Hoop}.   Faraday rotation from magnets are non-reciprocal whereas gyrotropic rotation is reciprocal.  Magneto-electrics can show both Faraday rotation as well as an interesting non-reciprocal phenomenon known as magneto-electric birefringence which manifests as different speeds of counter-propagating beams \cite{GB,OrensteinPRL11}.

A small polar Kerr effect on the order of $10^{-6}$ radian ($ \approx 10^{-4}$ degrees) has been recently detected using a high sensitivity Sagnac interferometer in the pseudogap phase of number of cuprate materials \cite{Kapi_YBCO,Kapi_instr, Kapi_LBCO2,Kapi_LBCO}.  In YBa$_2$Cu$_3$O$_y$ (YBCO), the effect onsets at temperature near, but systematically  below the temperature where a TRS breaking phase has been observed by neutron scattering \cite{Kapi_instr,Faque,Mook}.   In $x=1/8$  La$_{2-x}$Ba$_{x}$CuO$_4$ (LBCO)  the onset temperature coincides with the temperature where charge order develops.     As the Kerr rotation is constrained to be zero in states which break neither inversion or time-reversal symmetry the onset of a Kerr signal can be reasonably attributed to the breaking of one (or both) of these symmetries.  Initially, this effect was regarded as a magneto-optical, but later it was reported that the signal doesn't change sign with flipping the sample \cite{Kapi_LBCO,Kapi_LBCO2}, as expected in non-reciprocal magneto-optical effect, and a reciprocal gyrotropic charge order scenario was suggested\cite{Sri,oren_thy}.   Other models of non-reciprocal magneto-electric states and chiral states have been proposed to explain these unique features\cite{Varma, Victor}, but there is as of yet no agreed upon explanation for this effect.

Searching for such effects in a different frequency range may reveal essential information, since different physical mechanisms have different frequency dependencies. Magneto-optical effects may become enhanced at lower frequencies.   Gyrotropic effects should have a leading dependence that goes as $\omega \gamma(\omega)$ where $\gamma(\omega)$ is the $gyrotropic$ $parameter$.   The effect should then vanish in the long-wavelength limit, but  $\gamma(\omega)$ itself can have a complicated non-trivial frequency dependence the exact form of which is model dependent \cite{Sri, oren_thy}.   An interest in characterizing the frequency dependence of these effects is the main motivation for performing THz range polarimetry on the cuprates.  The typical probing energies of our experimental setup are almost three orders of magnitude lower than the near-IR frequency used in the Sagnac interferometer and more closely match the natural frequency scales of the pseudogap regime (1 THz $\approx$ 48 K).

Using a recently developed ``fast polarizer rotator" technique with time-domain THz spectroscopy (TDTS) \cite{Chris}, we were able to determine changes in the polarization of the transmitted light through high-T$_c$ superconductors thin films to high precision.  To within our experimental uncertainty (1.3 milli-radian or 0.07$^\circ$), no Faraday effect was found in this frequency range.   However in both the YBCO and LBCO systems a small, but significant anomalous linear birefringence/dichroism was found.  Its temperature onset mirrors the pseudogap temperature T$^*$, is enhanced in magnitude in underdoped samples, and shows a close correspondence to  the temperature dependencies found in the Kerr rotation.   It can can be associated with the loss of both C$_4$ rotation symmetry and mirror symmetry of the CuO$_2$ planes in the pseudogap state.

\begin{figure} 
\begin{center}
\includegraphics[width=7cm]{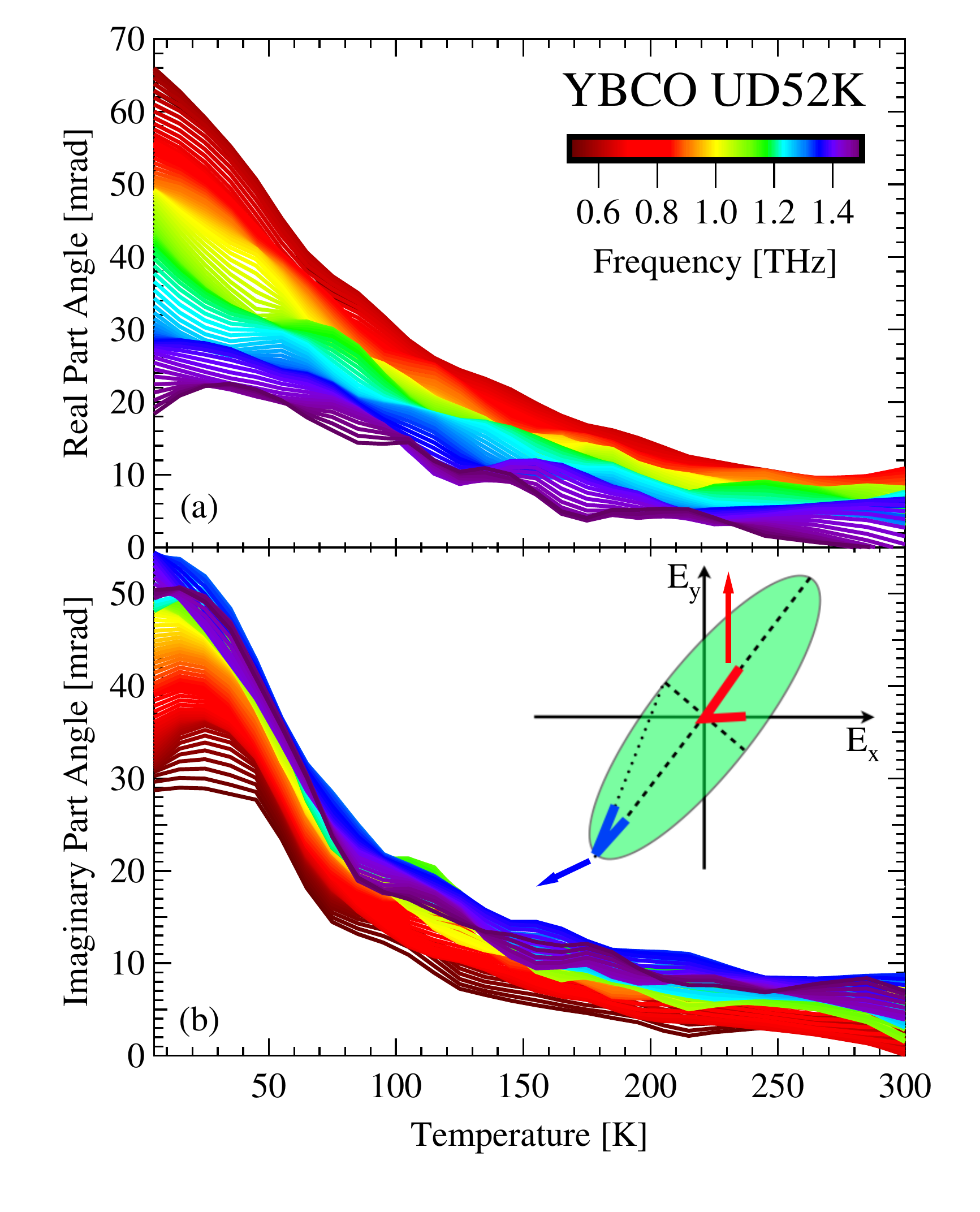}
\end{center}
\vspace{-1cm}
\caption{ The polarization rotating angle in transmitted THz beam through under-doped YBCO thin film (T$_c$=52 K) as a function of temperature and frequency. (a) The real part which represents the polar rotating angle. (b) The imaginary part which represents the ellipticity. The inset diagram demonstrates each quantity in a tilted oval propagating wave.  }
\label{Fig1}
\end{figure}

Several YBCO thin films and a LBCO one were grown by PLD on SrLaAlO$_4$ [001] tetragonal substrates.  The YBCO films spanned different doping levels from under-doped to over-doped.  In the YBCO films doping was controlled by an annealing process which affects the oxygen concentrations.   LBCO was grown with an $x=1/8$  target.  In TDTS one measures the transmitted $E$ field of a transmitted $\sim$ps long pulse as a function of time.   In usual TDTS one takes the ratio of Fourier transforms of the time-domain signals transmitted through the sample and a reference to get a $complex$ transmission function.  In the present work a THz pulse linearly polarized along an axis $x'$ propagates through a sample and we take the ratio of the complex transmitted electric field along the orthogonal direction $y'$ to the transmitted field along $x'$.    The arctan of this quantity defines a complex effective rotation angle.  After repeating the same procedure on an identical bare substrate as a reference, the rotation angle of the sample $\theta$, could be extracted by the subtraction of the two angles with a sensitivity generally better than 0.5 milli-radian (0.03$^\circ$).  The bare substrate was measured independently and didn't exhibit any notable optical response.  The real part of the complex angle $\theta$ is the angle of rotation of the major axes of the elliptically polarized transmitted light (related to dichroism) and the imaginary part is the ellipticity itself (inset to Fig. \ref{Fig1}.) \cite{DataExp} (related to birefringence).   In the present case \cite{DataExp}, the complex angle simplifies to the ratio $\sigma_{y',x'} / \sigma$ where $\sigma_{y',x'}$ is the off-diagonal conductivity in the reference frame where the incoming electric field is directed along $x'$ and $\sigma$ is the average diagonal conductivity.

\begin{figure}
\begin{center}
\includegraphics[width=7.5cm]{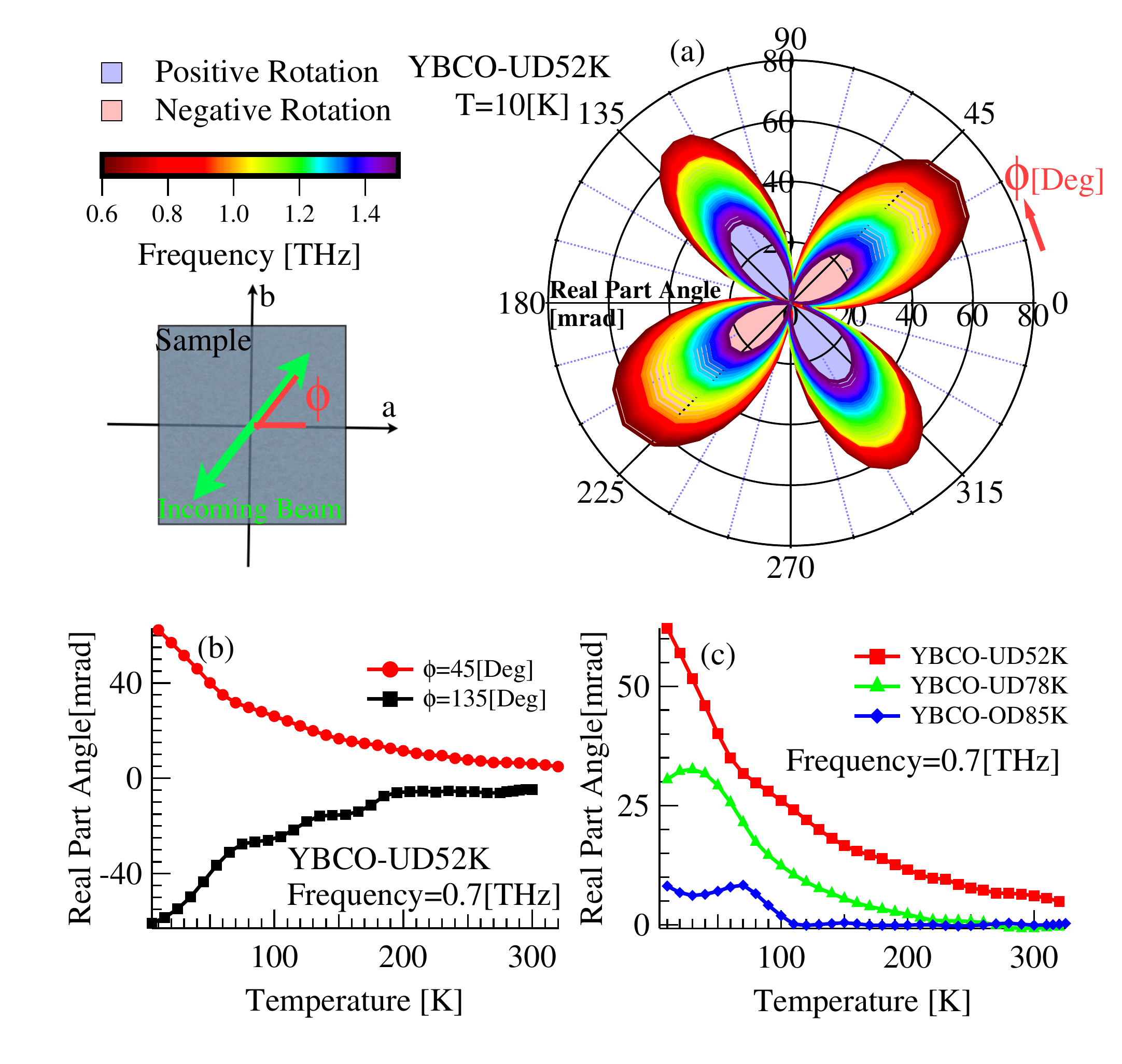}
\end{center}
\vspace{-1cm}
\caption{
The rotation angle as a function of the film's orientation $\phi$.  The panel in the top left demonstrate the definition of $\phi$ as the angle between the incident polarization and the film's lattice constants. (a) Polar presentation of rotating angle (radial axis) vs. $\phi$ (angular axis) in under-doped YBCO (T$_c$=52 K) at T=10 K. The direction of the rotation changes periodically from positive (blue background) to negative (red background). (b) Temperature dependence of the effect in two orthogonal directions from the same film.  (c) Real part of the rotating angle vs. temperature of three different doping.}
\label{Fig2}
\end{figure}

Typical data for the effective rotation angle in an underdoped ($T_c = 52$ K) YBCO sample as a function of temperature is shown in Fig. \ref{Fig1}.   The incident THz pulse's electric field angle $\phi$ was at 45$^\circ$ to the Cu-O bonds.  The signal increases as the temperature decreases in both real and imaginary parts with an onset temperature above room temperature.  Since the real part angle is the quantity reported in the near-IR previous work, we will focus on it, but it is important to keep in mind that the imaginary part provides important information as well.  Before comparing this rotation to the near-IR Kerr effect, we should verify whether this effect originates in a circular optical activity, like the one measured by the Sagnac technique. This can be done by rotating the incident polarization (or alternatively the sample) and mapping the rotation angle as a function of the film's orientation (represented by $\phi$ which is the angle between the incident wave's polarization and the lattice axes, as demonstrated in Fig. \ref{Fig2}).   For pure circular activity, a constant value of polarization rotation should be detected throughout all orientations.


A complete mapping of the signal as a function of the incoming electric field's angle $\phi$ is shown in Fig. \ref{Fig2}(a). The signal is clearly periodic, switching from rotating in one direction (positive) to the other (negative)  approximately every 90 degrees. The shape of this function deviates slightly from a perfect sine and has a slightly  sawtooth appearance, which is characteristic of linear dichroism, meaning absorption anisotropy in the plane.   With linear dichroism an effective rotation can be introduced when the incidence plane of polarization is not along an axis of symmetry.   This anisotropy has a distinct doping dependence, with both the size of the effect and its onset temperature becoming enhanced as the doping decreases as shown in  Fig. \ref{Fig2}(c).

We believe that this anisotropy represents an intrinsic electronic anisotropy of the CuO$_2$ planes in YBCO.  This is supported by a number of related facts.  Although the orthorhombic crystal structure of YBCO can in principle give linear dichroism, the doping dependence is $opposite$ to what one would expect in that scenario as underdoping YBCO moves it towards the tetragonal phase.   Moreover the Cu-O chains in YBCO become less conductive in underdoped samples further decreasing the possibility for structural orthorhombicity to give this effect.  We note that in all cases the onset temperature in YBCO is very close to the pseudogap's crossover temperature, T$^*$ (defined by DC transport measurement).   Note also that the linear dichroism's pattern is also anomalous in that it is tilted slightly (9$^\circ$ for the most underdoped YBCO) with respect to the crystal axes of the material.  This tilt is larger in underdoped films.

The existence of an effect dependent on linear polarization states does not necessarily mean the absence  of a circular one.  In the present experiment it would manifest as an offset in the periodic pattern, giving an asymmetry in the maximum positive and negative rotation and causing the positive and negative lobes to be different sizes.  Considering all the measurements with different films,  different doping, different families and different temperatures no such offset was detected. We are able to set an upper limit of 1.3 milli-radian (0.07$^\circ$) in which no circular effect is found above it in the THz regime.  We cannot of course rule out smaller effects below our detection sensitivity.  Recent models predict the possibility of a small circular effect with a corresponding large linear one in bilayers compounds, which could be appropriate in materials like YBCO\cite{Victor}. On the other hand,  similar linear behavior was observed also in monolayer LBCO as shown below.

\begin{figure}
\begin{center}
\includegraphics[width=7.5cm]{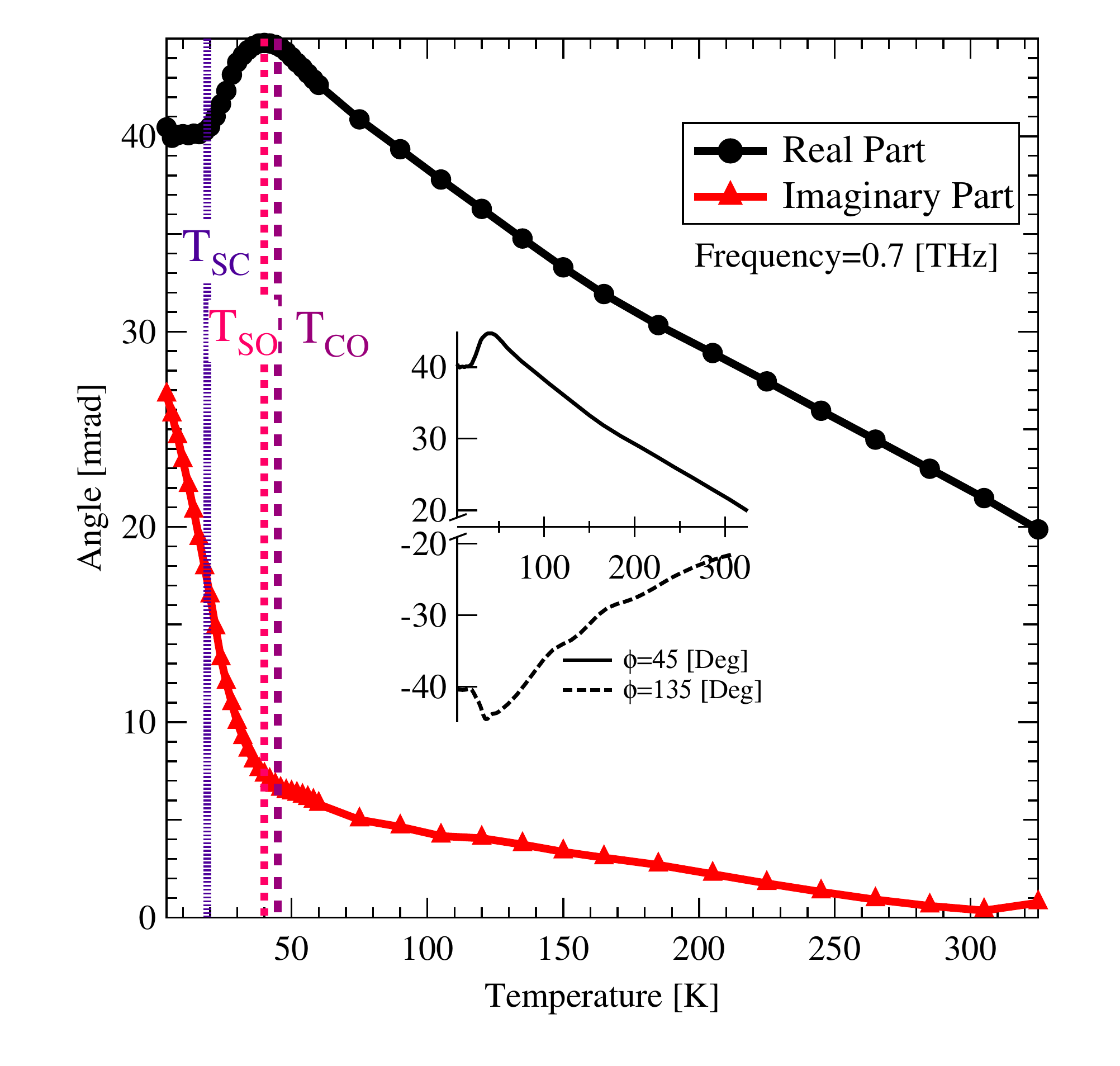}
\end{center}
\vspace{-1cm}
\caption{Real and imaginary parts of the rotation angle vs. temperature of the LBCO film.  Inset:  Temperature dependence of the effect in two orthogonal directions from the same LBCO film.  }
\label{Fig3}
\end{figure}

It interesting to compare our findings in YBCO with similar features we have found in LBCO.   LBCO $x=1/8$ single crystals have transitions at  T$_{HT}\approx$230 K\cite{lbco} from a high-temperature tetragonal (HTT) to a low-temperature orthorhombic (LTO) state and at T$_{LT}\approx$50 K to a low-temperature tetragonal (LTT) state.  However it is believed that in the case of films, the compressive strain from the LSAO substrate suppresses LBCO's LTT state and leaves the films in HTT or HTO states at low-temperature \cite{Sato00a}.  This can help distinguishing between trivial structural sources of the birefringence and more exotic phenomena.  In Fig. \ref{Fig3}. we show results of similar polarization rotation experiments on $x=1/8$ LBCO with incoming electric field at 45$^\circ$ to the Cu-O bond.   In the room temperature $tetragonal$ state there is already a significant effect in the real part of the rotation angle while the imaginary part is small.  By extrapolation of the temperature dependence, we can roughly estimate the onset temperature of the real part as T$_{onset}$=550(20) K, which is in rough agreement with the reported T$^*$   of this compound\cite{NMR}.   Both quantities become enhanced as the temperature is lowered with a distinctive temperature dependence.  Rotating the film by 90$^\circ$ (Fig. \ref{Fig3} inset) shows that the dependence flips sign confirming its origin in a linear dichroism as in YBCO.   Although it is believed that the compressive strain from the LSAO substrate suppresses LBCO's transition to the LTT state, \cite{Sato00a} it is important to check if any remnants of other transitions remain.  One might expect that such structural changes of LBCO would cause kinks in the linear birefringence which would appear in the imaginary part of the angle (Fig. \ref{Fig3}). In bulk $x=1/8$ LBCO the first structural transition occurs at  T$_{HT}\approx$230 K from HTT to LTO, but no sign of it is observed in the temperature dependence.   In bulk single crystals, the lower transition from LTO to LTT is T$_{LT}\approx$55 K.  We do observe a sudden enhancement of the imaginary part of the angle at a  lower temperature T$\approx$45 K, but the anisotropy enhancement is opposite to any expectation in the tetragonal phase.  This observation and that we see no sign of the higher temperature transition leads us to the conclusion that no direct signature of structural changes are detected, and that the effects we observe originate in the electronic structure.  We associate the sudden changes in the real and imaginary parts T$\approx$45 K with the spin- or charge-order transitions found at these temperatures in crystals.  

\begin{figure}
\begin{center}
\begin{tabular}{cc}
\includegraphics[width=4.3cm]{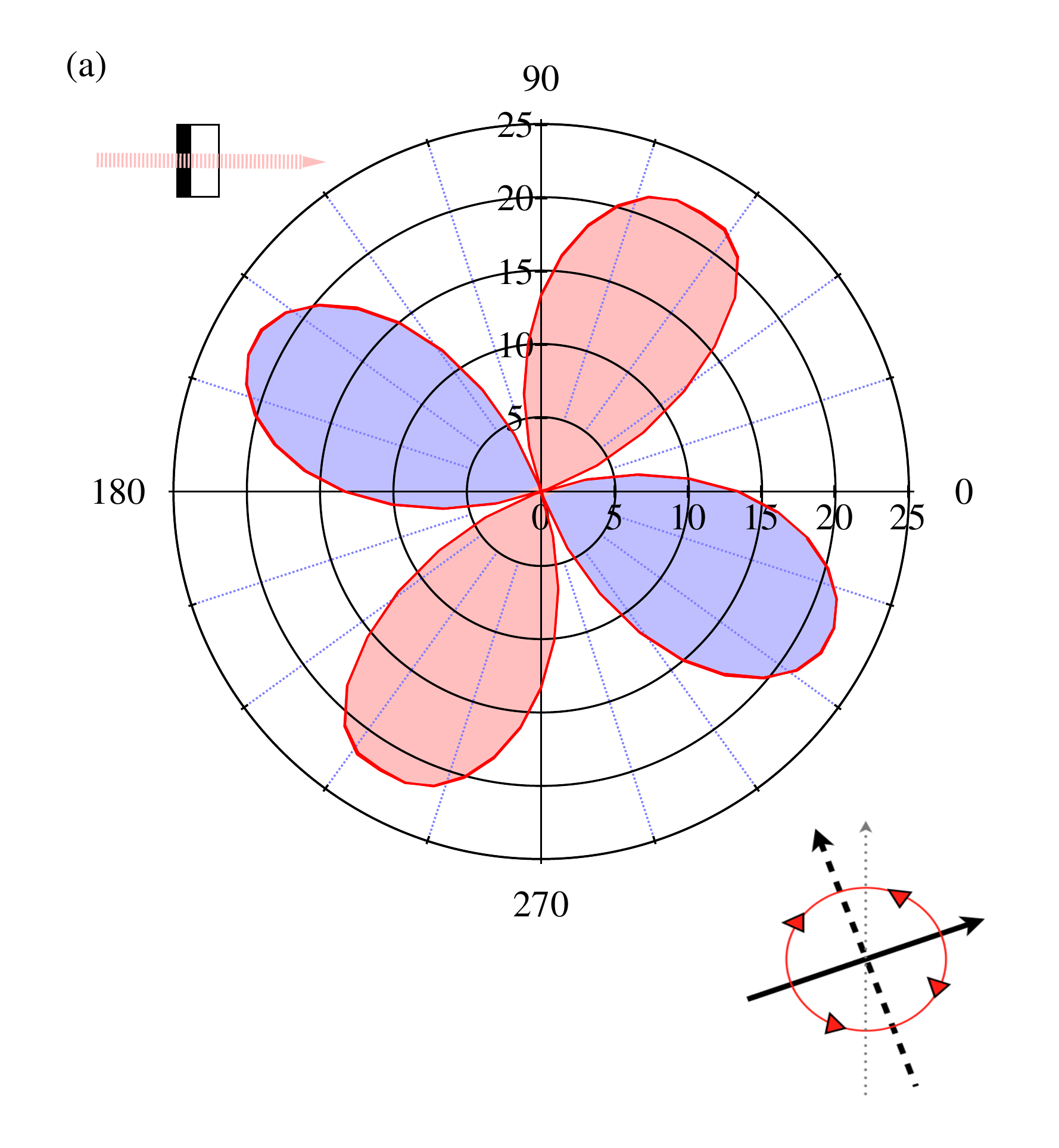}&
\includegraphics[width=4.1cm]{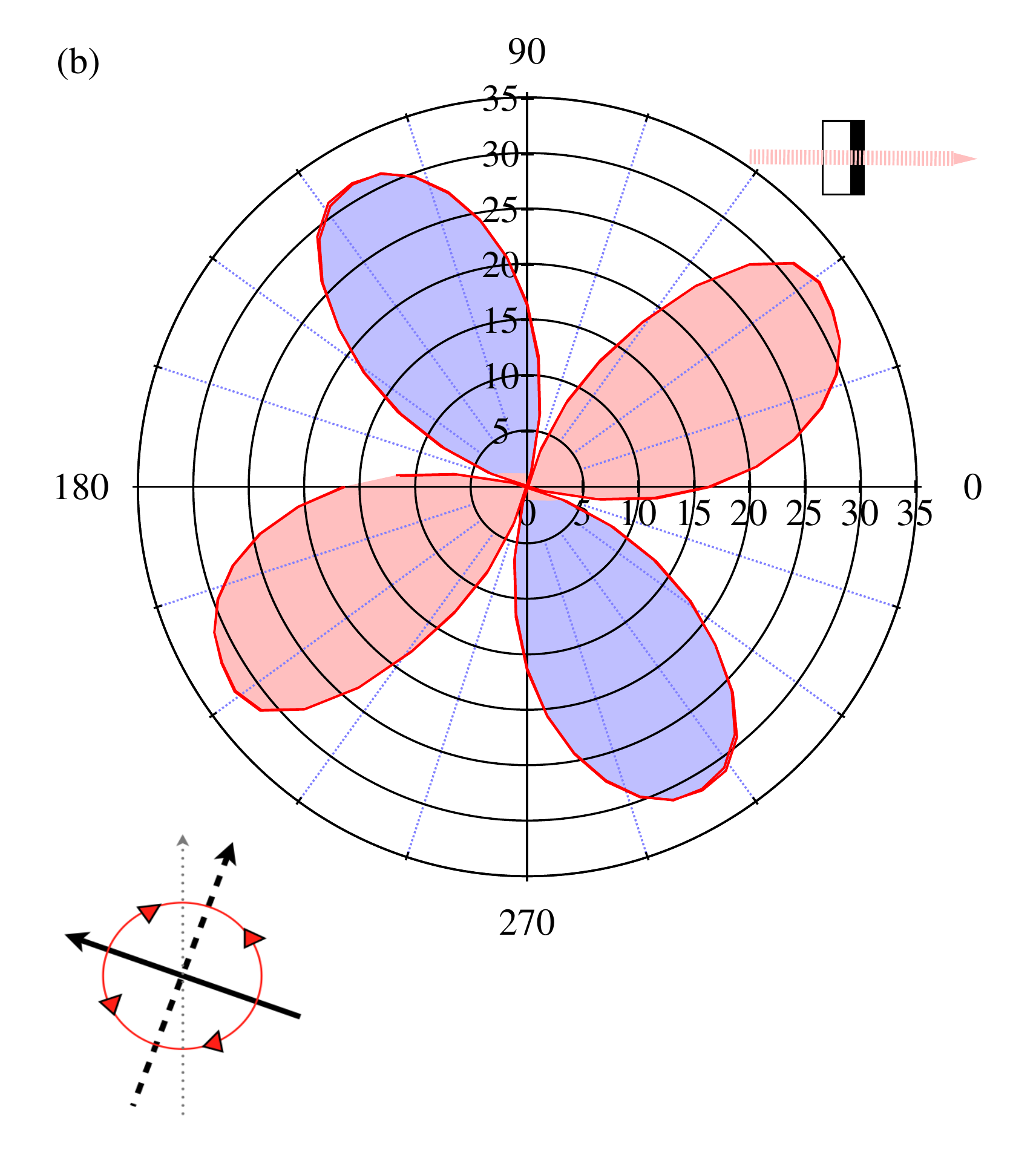}
\end{tabular}
\end{center}
\vspace{-1cm}
\caption{The quantities plotted are as explained in Fig. \ref{Fig2}(a). (a) and (b) The dichroism patten under flipping the film at T=300 K. The upper cartoon demonstrate forward and reverse propagation through the film and the substrate. The lower diagram represents a schematic of flipping the sample. The black arrows  represent the optical axes. The dotted gray arrow shows the flipping axis.  The red ones indicate the direction of the polarization rotation. We believe that the small difference in magnitude about the size of the effect for forward and reverse propagation is due to the slightly different effect of focusing in the substrate in the two directions.}
\label{Fig4}
\end{figure}

As shown in Fig. \ref{Fig4}, this rotation again derives from a linear polarization anisotropy and displays the distinctive four lobe pattern, with again the anomalous twist in the fact that the zeros of birefringence are rotated  (14(2)$^\circ$ in case for LBCO) with respect to the crystalline ones.  The pattern of birefringence  (Fig. \ref{Fig4}) flips when the sample is rotated around the (010) axis of the substrate and illumination is from the ``backside".  The changes are consistent  with normal linear birefringence flipped along an axis which is not a symmetry axis.  In the present case however, the flip was along a symmetry direction of either the tetragonal or orthorhombic axes.  We have verified through Laue diffraction that the crystal axes of the substrates are aligned with the substrates macroscopic edges to within $0.5^\circ$.   In the case of YBCO Raman microscopy has confirmed that no substantial regions of the samples have an in-plane (110) orientation (not shown).

We have shown evidence for an electronic structure anisotropy in the CuO$_2$ planes.  In accord with the discussion in the introduction, we interpret this as loss of a $C_4$ rotation symmetry of the electronic structure.   The tilt of the pattern of dichroism shows the loss of mirror symmetry.   It is interesting to point out that the systematics of the temperature dependence bears strong similarity to results on the Kerr rotation.  We believe these effects are related.   In YBCO, magnitude of this symmetry breaking and its temperature onset mimic the known dependencies of the pseudogap state while having an opposite dependence to the known orthorhombicity (although the onset temperatures are higher and close to the T$^*$ line in the present case).    In the case of LBCO our signal is finite at room temperature, but shows an enhancement near a temperature scale known to be associated with spin and charge ordered states (like in the Kerr rotation), a peak 10K below that, and flattening out of the real part in the superconducting state.  How to understand these results?

The linear dichroism we observe may suggest a tendency towards charge ordered structures with a 1D character.    Previous optical measurements in the cuprates have observed anisotropy that were interpreted in terms of stripe correlations \cite{basov, gedik, oren}.   We note that previously anisotropies in YBCO's DC resistivity had been observed that were attributed to these effects \cite{Ando02a}.  In the case of YBCO this anisotropy was reported to be in directional registry with the orthorhombic crystal lattice directions (the experiments were not done to sufficient precision to look for ``tilt").  Like in our results, it was believed to be associated with an electronic anisotropy that is not primarily dependent on structure, because it increased in magnitude with underdoping on the approach to the tetragonal phase.  However, it is not straightforward to understand how a macroscopic anisotropy would manifest itself in disordered, twinned and multi-domain thin films like the ones used in this work.  In the absence of some aligning field, conventional stripes structures should not keep the same orientation on a macroscopic scale and the corresponding optical effect is expected to be averaged out.  It is possible that features in the substrates (e.g. step edges) or mesoscopic structures in the samples (e.g. twin boundaries) could provide the aligning field.  We point out that similar issues with the lack of averaging over a domain structure must be confronted in the interpretation of the Kerr rotation experiments \cite{Kapi_LBCO2} which see a single helicity on both crystal faces and when probing length scales down to 10 microns.

An interesting scenario which may account for the unique tilted optical axes is gyrotropic magneto-electric birefringence. In this nonreciprocal magneto-electric effect, certain magnetic crystal structures that break both time and inversion symmetries can -- depending on symmetry and light propagation direction -- cause rotations and/or a linear birefringence with its zeros not along crystalline axes directions \cite{GB,OrensteinPRL11}.   It has been investigated most extensively for the case of Cr$_2$O$_3$ \cite{Pisarev91}. In principle this scenario can account for a number of unique features of our data including the existence of the tilt.  A recent model based on a orbital currents, predicts a magneto-electric effect that can account for aspects of our work (including a ``tilt" angle), has no sensitivity to domain structure, and similar to what is observed has -- at a lower temperature -- a transition to another loop-current state which breaks mirror symmetry resulting in a finite Kerr rotation \cite{Varma,VarmaLDichroism}.   It remains to be understood however, if models of this class can quantitatively account for the tilt angle we observe. 

We have demonstrated a loss of mirror and $C_4$ rotation symmetries in the pseudogap state that are not directly related to the structural features present in YBCO and LBCO materials.  Going forward it would be particularly interesting to perform similar  measurements in a reflection geometry on single crystals and to try extend these measurements to higher frequencies, particularly to the range where the Kerr effect is observed.

We would like to thank C. Naco and N. Drichko for performing the Raman characterization.  We  acknowledge useful discussions with J. Cerne, D. Golubchik,  A. Kapitulnik, S. A. Kivelson, J. Orenstein, S. Raghu,  K.M. Shen, J. Tranquada, C. M. Varma, J. Xia, and V. M. Yakovenko.  This THz instrumentation development and measurements at JHU was supported by the Gordon and Betty Moore Foundation through Grant GBMF2628 to NPA.  The film growth at Technion was supported in part by the joint German-Israeli DIP project and the Karl Stoll Chair in Advanced Materials.


\begin{thebibliography}{99}
\bibitem{DataExp}
  For a film of thickness $t$ on a substrate with index of refraction $n$ and for the case where both the off-diagonal conductivity tensor matrix elements and the differences in diagonal elements are small, the complex transmission angle is related to the elements of the complex conductivity tensor as $\theta \approx \sigma_{y',x'} t Z_0 / (1 +n +  \sigma t Z_0)$.   Here $\sigma_{y',x'}$ is the off-diagonal conductivity in the reference frame where the incoming electric field is directed along $x'$, $\sigma$ is the average diagonal conductivity, and $Z_0$ the impedance of free space  (377 $\Omega$). One can see that in the high conductivity limit, the complex angle is equal to the ratio $\sigma_{y',x'} / \sigma$.
\end{thebibliography}
\end{document}